\documentstyle[12pt]{article}
\newcommand{\la}{\label}

\newcommand{\X}{{\cal X}}

\newcommand{\vecn}{{\bf \vec n}}
\newcommand{\vece}{{\bf \vec e}}
\newcommand{\vecu}{{\bf \vec u}}

\newcommand{\be}{\begin{equation}}
\newcommand{\ee}{\end{equation}}
\newcommand{\ba}{\begin{eqnarray}}
\newcommand{\ea}{\end{eqnarray}}
\newcommand{\bastar}{\begin{eqnarray*}}
\newcommand{\eastar}{\end{eqnarray*}}

\begin{document}
\begin{titlepage}
\vskip 0.4truecm
\begin{center}
{ \bf \large \bf CHIRALITY AND FERMION NUMBER \\ \vskip 0.3cm
IN A KNOTTED SOLITON BACKGROUND \\ \vskip 0.3cm
}
\end{center}

\vskip 1.0cm
\begin{center}
{\bf Lisa Freyhult and Antti J. Niemi } \\

\vskip 0.3cm

{\it Department of Theoretical Physics,
Uppsala University \\
P.O. Box 803, S-75108, Uppsala, Sweden }\\

\end{center}

\vskip 4.5cm

\rm
\noindent
We consider the coupling of a single Dirac fermion 
to the three component unit vector field which appears
as an order parameter in the Faddeev model. Classically, 
the coupling is determined by requiring that it preserves 
a certain local frame independence. But quantum 
mechanically the separate left and right chiral fermion 
number currents suffer from a frame anomaly. We employ this 
anomaly to compute the fermion number of a knotted soliton. The
result coincides with the self-linking number of the soliton.
In particular, the anomaly structure of the fermions
relates directly to the inherent chiral properties of the soliton. 
Our result suggests that interactions between fermions 
and knotted solitons can lead to phenomena akin the Callan-Rubakov 
effect.
\noindent\vfill

\begin{flushleft}
\rule{5.1 in}{.007 in} \\
{\small Supported by NFR Grant F-AA/FU 06821-308} \\
{\small  E-mail: \scriptsize
FREYHULT@TEORFYS.UU.SE, 
NIEMI@TEORFYS.UU.SE}
\\
\end{flushleft}
\end{titlepage}

Knotted solitons \cite{nature} are present in a variety of field
theory models that appear both in high energy physics \cite{su2}
and condensed matter physics \cite{egor}. A canonical example of
a Lagrangian that describes knotted solitons is
the relativistic Faddeev model \cite{ludvig}
\be
L \ = \ (\partial_\mu \vecn )^2
\ - \ \Lambda ( \vecn \cdot \partial_\mu 
\vecn \times \partial_\nu \vecn)^2
\la{fada}
\ee
Here $n^a$ is a three component vector field, with
$\vecn \cdot \vecn = 1$. For a static finite energy 
configuration $n^a$ determines a mapping $R^3 \sim S^3 
\to S^2$. Such mappings are classified by $\pi_3(S^2) 
\sim Z$ which is computed by the Hopf invariant of $n^a$. 
The Hopf invariant of $n^a$ also coincides with the self-linking 
number of the knotted soliton, it can 
be either positive or negative reflecting the 
chirality of the knot \cite{nature}.

In \cite{su2} it has been proposed
that the Lagrangian (\ref{fada}) approximates the infrared 
limit of the four dimensional SU(2) Yang-Mills theory. 
Since a Yang-Mills theory usually couples to 
fermions, it should be interesting to consider also
fermions in interaction with (\ref{fada}). Furthermore,
the fermion number current of a knotted soliton is 
an important quantity in condensed matter physics 
where it couples to the Maxwellian electromagnetic
field. 

In the present Letter we shall compute 
the fermion number \cite{gordon} of a knotted soliton in 
the Faddeev model (\ref{fada}). For this
we need to determine how $n^a$ couples to  
a fermionic field. Previously such couplings have 
been considered in the context of supersymmetric O(3) 
nonlinear sigma-models. There, the coupling
between a three component vector field $n^a$ and its 
supersymmetry partner, a three component fermion 
$\psi^a$, is commonly introduced by imposing the condition 
$n^a \psi^a = 0$; See {\it e.g.} \cite{berg}. 
Here we are interested in the 
coupling of $n^a$ to a {\it single} Dirac fermion
with no supersymmetry. For this we introduce
an explicit realization of $n^a$, viewing it 
as a unit normal to a sphere $S_n^2 \in R^3$,
\be
n^a \ = \ \pmatrix{ \cos \phi \sin \theta \cr 
\sin\phi \sin \theta \cr \cos \theta
} \ = \ {\cal O}^{a}_{b}u^b_3 \ = \ 
\pmatrix{ \cos \phi \cos \theta & - \sin \phi & \cos \phi
\sin \theta \cr \sin \phi \cos \theta & \
\cos \phi & \sin \phi \sin \theta
\cr - \sin \theta & 0 & \cos \theta } \pmatrix{ 0 \cr
0 \cr 1 }
\la{na}
\ee
where we interpret $\vecn$ as a vector which 
is obtained by rotating the unit vector
$\vecu_3 = (0,0,1)$ with the SO(3) matrix 
${\cal O}^a_b$. We also introduce two unit vectors 
$\vece_{1}$ and $\vece_2$ which are tangent to 
the sphere $S_n^2$. These 
vectors can be obtained similarly, by rotating the unit 
vectors $\vecu_1=(1,0,0)$ and $\vecu_2=(0,1,0)$ 
with the matrix ${\cal O}^a_b$,
\be
\vece_{\pm} \ = \ \frac{1}{\sqrt{2}}( \vece_1 \pm i \vece_2) 
\ = \ \frac{1}{\sqrt{2}} 
\pmatrix{ \cos \phi \cos \theta 
\mp i \sin \phi \cr \sin \phi \cos 
\theta \pm i \cos \phi \cr - \sin 
\theta} e^{\pm i \X} \ \equiv \ V_{\pm}
\la{epm}
\ee
Here $\X$ is an angle that specifies the orientation
of the combinations
$\vece_{\pm}$ with respect to the 
${\cal O}^a_b$ rotation of the
canonical basis vectors
$\vecu_{1,2}$. {\it A priori} there is no reason 
to select $\X = 0$. Instead, this variable determines 
an inherent degree of freedom that we have in 
defining an orthonormal basis on the tangent plane 
of $S_n^2$. 

The last relation in (\ref{epm}) interprets $\vece_{\pm}$ 
as $3 \times 1$ complex matrices $V_{\pm}$, 
with $\vece_\pm$ as the single vertical matrix column. 
These matrices have the properties 
$V^{\dag}_{\pm} V_{\pm} = 1$ and
$V_{\pm} V^{\dag}_{\pm} = P_{\pm}$ 
with $P_{\pm}$ the projection operators 
to the chiral subspaces which are spanned by $\vece_{\pm}$. 
Consequently we can employ $V_{\pm}$ to 
define two (chiral) spin connections $A^{\pm}_\mu$ \cite{nara},
\be
A^{\pm}_\mu dx^\mu \ = \ V_{\pm}^{\dagger} d V_{\pm} 
\ = \ \mp i(\cos \theta d\phi - d\X)
\la{rama}
\ee
Together with the ensuing curvatures
\be
F^{\pm} \ = \ dV^{\dag}_{\pm} (1-P_{\pm})
dV_{\pm} \ = \ \pm i \sin \theta d\theta \wedge d\phi 
\ = \ \pm \frac{i}{2} \vecn \cdot d\vecn \wedge d\vecn 
\la{fmunu}
\ee
we obtain for the Hopf invariant of $n^a$ 
\be
Q_H \ = \ \frac{1}{8\pi^2} \int A^{\pm} \wedge F^{\pm} \ = 
\ \frac{1}{8\pi^2}
\int \sin \theta d\theta \wedge d\phi \wedge d\X
\la{hopf1}
\ee
which computes the self-linking number of the knotted soliton 
$n^a$. We note that (\ref{hopf1}) relates the 
angular variable $\X$ in (\ref{epm}) to a coordinate variable for 
the arc-length along the knotted soliton. 
Clearly, $\X$ must be nontrivial whenever the Hopf-invariant is
nonvanishing {\it i.e.} whenever $n^a$ forms a knot or a link. 
We also note that the Hopf invariant can be either positive 
or negative. This reflects the inherent chiral structure 
of knotted solitons, see \cite{hie} for a detailed description.

Consider a standard four-dimensional Dirac fermion $\psi_D$.
Since the knotted solitons have a chiral
structure which is reflected by (\ref{rama}), we
also decompose $\psi_D$ into its right {\it resp.} 
left chiral Weyl components, 
$$
\psi_D = \pmatrix{\psi_+ \cr \psi_-}
$$
with $\gamma^5 \psi_{\pm} = \pm \psi_{\pm}$.
We then couple $\psi_D$ and $n^a$ by minimally
coupling the chiral components $\psi_{\pm}$ to the ensuing 
spin connections $A^{\pm}_\mu$. The interaction Lagrangian
is
\be
L_{int} \ = \ ig_+ A^+_\mu \psi^{\dagger}_+ \bar\sigma^\mu \psi_+
\ - \ ig_- A^-_\mu \psi^{\dagger}_{-} \sigma^\mu \psi_- \ = \
ig_+ A_\mu j^+_\mu - ig_- A_\mu j^-_\mu
\la{coupl}
\ee
where $A_\mu = A_\mu^+ = - A_\mu^-$ and 
$g_{\pm}$ are dimensionless
coupling constants that we shall specify
shortly. Clearly, this is the simplest natural  
coupling of $\psi_D$ with the degrees of freedom in $n^a$ 
which is manifestly invariant under global SO(3) 
rotations of $n^a$. Furthermore, we note
that at the classical level 
the coupling is also invariant under local rotations of 
the $\vece_{\pm}$ frame, when compensated by a U(1) phase
redefinition of $\psi_{\pm}$. (Notice that the global SO(3) rotation 
invariance of $n^a$ must become broken when (\ref{fada}) 
describes the infrared limit of Yang-Mills theory. Here we assume that 
this symmetry breaking is extrinsic to fermions.) 

The fermion number of the knotted soliton 
is the space integral of the summed time component 
of the (properly normal ordered)
fermion currents in (\ref{coupl}) 
\be
N \ = \ \int <\! \bar \psi_D \gamma^0 \psi_D\!> \ = \ 
\int <\! j_0^+ + j_0^- \!> \ = \ N_+ + N_-
\la{fnum}
\ee
In order to compute (\ref{fnum}) we consider the
pertinent Dirac Hamiltonian in a representation 
of $\gamma$-matrices where $\gamma^0 = \sigma^1 
\otimes 1$, $\gamma^i = i \sigma^2 \otimes \sigma^i$ 
and $\gamma^5 = \sigma^3 \otimes 1$ so that
\be
H \ = \ \pmatrix { i \sigma^i (\partial_i + g_+A_i) & 0 \cr
0 & -i\sigma^i(\partial_i - g_- A_i) } 
\la{dirham}
\ee 
(Recall that $A_\mu = A^+_\mu = - A^-_\mu$.)
We note that at the quantum level 
the {\it separate} right and left
chiral components $j_\mu^\pm$ of the fermion number
current do suffer from an equal size but opposite
sign anomaly under the U(1) frame rotation 
\be
\vece_{\pm} \ \to \ e^{\pm i\varphi} \vece_{\pm}
\la{rot}
\ee
This anomaly is
\be
\partial_\mu j^\mu_+ \ = \ - \partial_\mu j^\mu_- \ = \ - 
\frac{1}{32\pi^2} \epsilon^{\mu\nu\rho\sigma} F_{\mu\nu}
F_{\rho\sigma} \ = \ -
\frac{1}{16\pi^2} \partial_\mu (\epsilon^{\mu\nu\rho\sigma}
A_\nu F_{\rho\sigma})
\la{anomaly}
\ee
But for consistency the coupling between $\psi_D$ and $n^a$ 
must conserve the fermion number current also at 
the quantum level. According to (\ref{rama}) this 
implies that we must set $g_+ = - g_-$ in (\ref{dirham}). 
The fermion number is then a conserved quantity also in the 
quantum theory.

The Hamiltonian (\ref{dirham}) and $\gamma^5$ commute, 
$ H \gamma^5 = \gamma^5 H$. Consequently we can choose all energy
eigenstates of $H$ to have a definite chirality. Furthermore,
since $ \gamma^0 \gamma^5 H = - H \gamma^0 \gamma^5$ the spectrum
of (\ref{dirham}) is symmetric which implies that the $\eta$-invariant
of $H$ vanishes. Thus the sole contribution to the
fermion number (\ref{fnum}) of a knotted soliton comes from a 
spectral flow in (\ref{dirham}) \cite{gordon}. For this we recall
the anomaly (\ref{anomaly}). With it, we compute 
the spectral flow contribution to the fermion number of 
a knotted soliton as follows: We first introduce an 
adiabatic process where we interpolate between the desired 
knotted soliton $n^a$ with a nontrivial Hopf invariant ({\ref{hopf1}), 
and another configuration with a vanishing Hopf invariant. 
The adiabatic interpolation is achieved {\it e.g.} by 
introducing a slowly varying function of time $f(t)$ such 
that $f( - \infty) = 0$ and $f(+ \infty) = 1$, and  
promoting the spin connections to $t$-dependent fields for 
example by defining 
\be
A_i^{\pm}(t)dx^i \ = \ \mp i ( \cos \theta d \phi - f(t) d\X)
\la{int1}
\ee
with the time components $A^\pm_0 = 0$. Here $\theta$, $\phi$ and $\X$ are 
$t$-independent fields that specify the desired static knotted 
soliton profile $n^a$. When $t \to +\infty$ 
(\ref{int1}) yields the spin connections that correspond to
the desired knotted soliton $n^a$ with its nontrivial Hopf invariant
(\ref{hopf1}). But for $t \to -\infty$ we have 
spin connections for which the Hopf invariant (\ref{hopf1}) 
vanishes; Obviously, for the adiabatic $t$-evolution the vector
$n^a$ can not be defined in a smooth manner. But 
the connection (\ref{int1}) and the Dirac Hamiltonian 
(\ref{dirham}) are well defined for all $t$ which is
quite sufficient for the present computation.
When we integrate the corresponding 
anomaly equations (\ref{anomaly}) 
over space-time the only contribution 
comes from $R^3$ at $t = +\infty$ which yields for the chiral 
fermion numbers in the background of 
the desired knotted soliton 
\be
N_+ = - N_- \ = \ -\frac{1}{16\pi^2} \int d^3x \epsilon^{ijk} A_i F_{jk}
\ = \ - Q_H
\la{diff}
\ee
so that during the adiabatic process we have $\Delta N_+ = 
- \Delta N_- = Q_H$. This means that the adiabatic formation of the 
knotted soliton is accompanied by a nontrivial spectral flow, 
where $|Q_H|$ positive (negative) chirality eigenvalues of $H$ 
cross from negative to positive values and an equal number
of oppositely chiral eigenvalues of $H$ cross 
symmetrically to the opposite direction. 

We assume that the initial Dirac vacuum at $t = -\infty$ is 
normal ordered canonically, with all of the $E<0$ energy 
levels of $H$ filled. When $t=+\infty$ we have then created
$|Q_H|$ physical chiral fermions, in
addition of the knotted soliton. Since the spectrum of $H$ 
is symmetric we have also formed $|Q_H|$ empty, oppositely 
chiral negative energy levels. These empty negative energy 
levels in the Dirac sea are then responsible for the fermion
number of the knotted soliton, which also coincides with its 
self-linking number. In particular, we conclude that 
during the entire process the {\it total} 
chirality and fermion number remain conserved.

More generally, the present considerations suggest interaction
processes where the initial states consist of an
incoming massless negative (or positive) chirality fermion 
in addition of a knotted soliton with self-linking number $Q_H$, 
and the final states consist of an oppositely charged, oppositely 
chiral fermion in addition of a knotted soliton with self-linking 
number $Q_H \pm 2$. Such processes could then lead to an excess 
in the number of chiral fermions {\it e.g.} in the Universe, 
together with oppositely chiral (knotted) solitons.

\vskip 0.8cm

In conclusion, we have coupled the Faddeev model to a 
single Dirac fermion and computed the ensuing fermion
number of a knotted soliton. The result coincides with
the self-linking number of the soliton, which is 
evaluated by the Hopf invariant. The computation employs the abelian 
chiral anomaly of the fermion, which we also find to be consistent with
the inherent chiral structure of the knotted soliton. 
Since Maxwellian electromagnetism couples directly to the fermion 
number current, our result suggests that in condensed matter 
physics applications the knotted solitons can be charged. 
In high energy physics applications where the Faddeev model 
supposedly describes the low energy limit of a pure Yang-Mills 
theory, our results can be used to inspect various properties
of knot-fermion interactions. Indeed, the non-vanishing of the 
fermion number of a knotted soliton and its intimate relation
with the fermionic chiral structure in fermion-knot interaction
processes provides a basis for interesting phenomenological scenarios, 
inclusive a version of the Callan-Rubakov effect.

\vskip 0.5cm
A.Niemi thanks Dmitri Diakonov for discussions and Nordita
for hospitality during this work.

\end{document}